\documentclass[aip,reprint,longbibliography]{revtex4-2}

\usepackage{amsmath,amssymb,bm}
\usepackage{graphicx}
\usepackage{physics}
\usepackage{siunitx}
\usepackage{hyperref}
\usepackage{color}

\begin{document}

\title{Application of the aperiodic defect model to a negatively charged monovacancy in phosphorene}

 \author{Charlotte Rickert}
 \affiliation{Institut f\"ur Chemie, Humboldt-Universit\"at zu Berlin, Brook-Taylor-Str. 2, Berlin 12489, Germany}
 \affiliation{Max Planck Institute for Solid State Research, Heisenbergstr. 1, 70569 Stuttgart, Germany}

 \author{Lily Barta}
 \affiliation{Institut f\"ur Chemie, Humboldt-Universit\"at zu Berlin, Brook-Taylor-Str. 2, Berlin 12489, Germany}

 \author{Ernst-Christian Flach}
 \affiliation{Institut f\"ur Chemie, Humboldt-Universit\"at zu Berlin, Brook-Taylor-Str. 2, Berlin 12489, Germany}

\author{Daniel Kats}%
\affiliation{Max Planck Institute for Solid State Research, Heisenbergstr. 1, 70569 Stuttgart, Germany}%

 \author{Denis Usvyat}
\email{denis.usvyat@hu-berlin.de}
 \affiliation{Institut f\"ur Chemie, Humboldt-Universit\"at zu Berlin, Brook-Taylor-Str. 2, Berlin 12489, Germany}

\date{\today}

\begin{abstract}
  We apply the recently introduced aperiodic defect model (ADM) to a negatively charged monovacancy in a phosphorene monolayer.  In contrast to conventional supercell approaches, the ADM treats a single defect embedded in the true non-defective crystalline mean field thereby avoiding spurious defect–defect interactions and the need for charge corrections. At the same time, it effectively reduces the calculation to a fragment, enabling the use of high-level molecular electronic-structure methods. Converging the Hartree-Fock and correlation contributions to the thermodynamic limit yields a benchmark CCSD(T)/POB-TZVP-rev2 formation energy of 0.81 eV for the negatively charged monovacancy in the (5$|$9) configuration. The excitation energy to the lowest singlet excited state of this defect at the EOM-CCSD/POB-TZVP-rev2 level is found to be 1.95~eV. Overall, the ADM provides a highly promising route towards quantitatively accurate and systematically improvable descriptions of defects in solids and on surfaces, bridging the gap between solid-state physics and molecular quantum chemistry. 
\end{abstract}

\maketitle

\section{Introduction}

Elemental phosphorus is an intriguing material for chemical and technological applications, particularly due to the diversity of its allotropes and their properties. Under standard conditions, black phosphorus has traditionally been considered the most thermodynamically stable form of phosphorus\cite{Jacobs_bl_P,corbridge2013phosphorus,Nilges18} (although recent studies suggest that violet phosphorus may be slightly more stable, yet within methodological uncertainty\cite{Vogt2023,Bonometti2025}).  Black phosphorus represents van-der-Waals-bound stacks of covalently-bonded monolayers.\cite{Hultgren35,Morita1986,ShulenburgerTomanek_QMC_BP,Sansone2016,schutz2017} An individual monolayer, referred to as phosphorene, is of considerable interest in its own right as a two-dimensional material with strong potential for applications in single-layer electronics,\cite{Li2014BPTransistor,Liu2014PhosphoreneTransistor,Zhang2024} optoelectronics,\cite{Xia2014RediscoveringBP,Zhang2024} catalysis,\cite{Jain_2017,Nehra01022024} and other domains.\cite{CastellanosGomez2015Review,Tian2023}

As in many low-dimensional materials, the properties of phosphorene can be strongly affected by the presence of point defects such as vacancies or substitutional impurities. In semiconductors in general, such defects may reduce charge carrier mobility by acting as scattering or trapping centers.\cite{Shockley1952,Stoneham2001} At the same time, the controlled incorporation of defects is essential for semiconductor technology, as defect-induced donor or acceptor states enable control of carrier concentrations.\cite{Sze2006,YuCardona2010} Beyond electronic transport, point defects can also modify chemical reactivity and catalytic activity by creating localized active sites.\cite{Xie2020_ACSCatalysis_Defects,Luo2023_Nanomaterials_Defects} One can also mention that certain defects can host localized electronic states relevant for quantum technologies.\cite{Doherty2013,Awschalom2018}

Point defects in phosphorene, in particular vacancies, have therefore attracted considerable theoretical attention.\cite{Wang2015NativeDefects,Guo2015VacancyStates,Srivastava2015VacanciesPhosphorene,HuYang2015_JPCCdefects,Hu2015_PhosphoreneVacancies,Carvalho2016Review,Cai2016,C9NR06608J,Naik2020_PRMaterials_PhosphoreneDefects,GONZALEZREYES2024416455}
Previous studies have shown that the monovacancy undergoes a substantial local reconstruction accompanied by rearranging of covalent bonds\cite{Hu2015_PhosphoreneVacancies,Wang2015NativeDefects} and introduces localized electronic states within the band gap.\cite{Guo2015VacancyStates,Srivastava2015VacanciesPhosphorene} At the same time, the formation energy for this defect has been found to be rather low compared to other 2D materials, such as graphene or silicene.\cite{HuYang2015_JPCCdefects} The migration barrier of the vacancy is also relatively low, suggesting that such defects may be highly mobile and dynamically active under suitable conditions.\cite{Cai2016,C9NR06608J}
The monovacancies in phosphorene have been predicted to act as acceptor-like defects and may contribute to the commonly observed p-type character of black phosphorus.\cite{Wang2015NativeDefects}

Most computational investigations of point defects employ density functional theory (DFT) within the periodic supercell approach.\cite{Freysoldt2014Review} While this approach has proven successful, it inherently introduces artificial periodicity of the defect causing non-physical interactions. The associated finite-size effects may compromise the description of isolated defects and slow down convergence to the thermodynamic limit. Particularly problematic are charged defects, whose periodization requires an unphysical compensating background.\cite{Freysoldt2009ChargeCorrection,Lany2008FiniteSizeErrors} Another challenge is description of defects beyond DFT. In contrast to DFT, wavefunction-based approaches, such as coupled cluster theory, offer a systematically improvable methodology, but for large supercells they quickly become computationally prohibitive. To address this issue, embedding strategies have been proposed that restrict the computationally expensive high-level correlated treatment to the defect region, while the environment is represented by a lower-level description.\cite{Govind1998,Kluner,Birkenheuer2006,Jacob08,Hozoi2009,de_lara-castells2011,bygrave2012,Manby2012,Libisch2014,Stoyanova2014,masur2016,Sun2016,Goodpaster2014,Jacob2014,Welborn2016,Libisch2017,usvyat18,Chulhai2018,Fertitta2018,Zhu2019,usvyat20,Lacombe2020,Cui2020,Jones2020,Pham2020,schaefer21,Hegely,Ma2021,christlmaier21,Berkelbach21,WachterLehn2022,Nusspickel2022}

In this work, we study the negatively charged monovacancy in phosphorene using the recently developed aperiodic defect model (ADM).\cite{Lavroff2025_JCP_ADM} In contrast to the supercell model, the ADM describes a single defect embedded in an otherwise non-defective crystal. It thus avoids the artificial periodicity of the supercell calculations and lifts the necessity for the compensating background charge and the associated corrections.  In the absence of the spurious long-range defect–defect interactions the convergence towards the thermodynamic limit is expected to be faster. Furthermore, since the ADM reduces the problem to an effective molecular-like calculation, it enables application of highly accurate molecular quantum-chemical methodologies for the description of both ground and excited states.

Nevertheless, a charged monovacancy in phosphorene is a challenging system also for ADM. Phosphorene forms a covalently bonded network that can undergo substantial structural rearrangement in the presence of a vacancy. In combination with the relatively high polarisability of this material, this may lead to a long-range response to the defect and consequently a slow convergence with respect to fragment size. In addition, the relatively narrow band gap may cause poorer orbital localization, which can obscure the fragment-environment partitioning. Finally, the present work considers ground and excited states of a charged defect, whose spatial locality and thus rapid convergence with fragment size is not guaranteed. For these reasons, we consider this application as an instructive test case for assessing the capabilities of the ADM.

\section{Aperiodic Defect Model}

The details of the aperiodic defect model can be found in Ref.~\onlinecite{Lavroff2025_JCP_ADM}, here we provide only a conceptual overview. In the ADM the first step is a periodic restricted Hartree-Fock (HF) calculation on a non-defective host crystal. The localized occupied orbitals --- Wannier functions (WFs) --- from this calculation\cite{zicovich-wilson2001,Zicovich_boys} are then used to partition the system into the environment and the fragment. The environment remains frozen in the bulk periodic HF solution, while the fragment's geometry is manipulated to create the defect.
The electronic structure of the fragment with the defect is then recalculated within an effectively molecular-like treatment, with the embedding field from the frozen environment implicitly included in the fragment's one-electron Hamiltonian 
\begin{eqnarray}
h^{\rm def}_{\mu'\nu'}&=&F^{\rm per}_{\mu'\nu'}-\sum_{i\in{\rm frag}}\left[2\left(\mu'\nu'|ii\right)
  - \left(\mu' i|i \nu'\right)\right]\nonumber\\
&-& \bra{\mu'} \sum_{K'\in{\rm add}}{Z_{K'} \over |{\bf r}-{\bf
    R}_{K'}|}-\sum_{K\in{\rm rem}}{Z_{K} \over |{\bf r}-{\bf
                          R}_{K}|}\ket{\nu'}.\label{eq:defect_h}
\end{eqnarray}
Here $F^{\rm per}$ is the Fock matrix from the non-defective periodic HF calculation, $(\mu'\nu'|ii)$ and $(\mu' i|i \nu')$ are the Coulomb and exchange integrals in the usual chemical notation, with $i\in{\rm frag}$ being the fragment-defining occupied WFs from the pristine HF calculation. The summations in eq.~(\ref{eq:defect_h}) over $K$ and $K'$ are carried out only over the manipulated atoms, i.e., the removed ones $(K)$ and the added ones $(K')$.

The atomic-orbital-like (AO-like) basis functions $\ket{\mu'}$ are obtained by projecting the AOs centered on the fragment atoms (after defect formation) out of the occupied Wannier-function manifold of the environment ($i \notin {\rm frag}$):
\begin{eqnarray}
  \left|\mu'\right>=\left(1-\sum_{i\notin{\rm frag}}\left|i\right>\left<i\right|
  \right)\left|\mu\right>.\label{eq:new_AO}
\end{eqnarray}
This projection guarantees orthogonality of the fragment orbital space to the occupied manifold of the frozen environment and at the same time allows the use of atom-centered basis functions with the same nomenclature as the usual AOs.

The fragment atoms that serve as centers for the basis orbitals (as well as auxiliary functions for density fitting) are chosen using the orbital-domain machinery of the periodic local MP2 code of \textsc{Cryscor}.\cite{pisani2008} The atomic fragment is thus defined as the union of the centers that enter the domains of the fragment-defining WFs.
In practice, because atomic fragments are easier to control and visualize, the procedure is reversed: the fragment is specified in the input in terms of atoms, while the fragment WFs are chosen such that their domains lie entirely within this atomic fragment. By default, the standard Boughton-Pulay domains\cite{Boughton1993} with a tolerance of 0.98 are used, which typically leads to one-atom domains for lone-pair WFs and two-atom domains for bonding WFs. The domains, and thus the set of AO centers for the fragment, can also be extended using, for example, distance-based or bond-connectivity criteria.

The number of electrons in the fragment is determined by the fragment WFs rather than by the formal charge of the fragment atoms:
\begin{eqnarray}
  N_{\rm el}^{\rm frag}=2N_{\rm WFs}^{\rm frag}-\sum_{K\in{\rm rem}}Z_{K}+\sum_{K'\in{\rm add}}Z_{K'}+\Delta_{\rm el},\label{eq:n_el}
\end{eqnarray}
where $N_{\rm WFs}^{\rm frag}$ is the number of initial WFs that define the fragment, $Z_{K}$ and $Z_{K'}$ are the nuclear charges of the removed and added atoms of the defect, respectively, and $\Delta_{\rm el}$ is the number of added or removed electrons if the defect is charged. A possible excess in the nuclear charge in the fragment is conceptually unproblematic, as it is compensated by the environment electrons.

The energy of the system is defined as the energy of the finite fragment subjected to the periodic embedding field. This leads to standard expressions for the HF or correlation energy formally equivalent to those of an isolated finite cluster, but with modified Hamiltonian matrix elements (\ref{eq:defect_h}) and an effective nuclear energy:
\begin{eqnarray}
  E^{\rm def}_{\rm nuc}&=&{1 \over 2}\sum_{K'\in{\rm add}}\sum_{L'\in{\rm add}}\phantom{}^{'}{Z_{K'}Z_{L'} \over |{\bf R}_{K'}-{\bf
      R}_{L'}|}\nonumber\\
  &&+\sum_{K'\in{\rm add}}\left[ Z_{K'}\cdot V\left({\bf
                          R}_{K'}\right) -\sum_{K\in{\rm rem}}\phantom{}^{'}{Z_{K}Z_{K'} \over |{\bf R}_{K}-{\bf R}_{K'}|}\right]\nonumber\\
  &&-2\sum_{i\in{\rm frag}}\left< i\left|-\sum_{K'\in{\rm add}}{Z_{K'} \over |{\bf r}-{\bf  R}_{K'}|}\right|i\right>,\label{eq:E_nuc_defect}
  \end{eqnarray}
where $V\left({\bf R}_{K'}\right)$ is the electrostatic potential from the periodic host crystal at the position of the added atoms. 
Again the summations over nuclei run only over the removed or added ones rather than over the entire fragment. This restriction only affects a constant energy shift which cancels out in energy differences.\cite{Lavroff2025_JCP_ADM}

The density-fitted ADM has been implemented in the \textsc{Cryscor} program\cite{usvyat18,Lavroff2025_JCP_ADM} by repurposing the local density-fitting machinery of the periodic LMP2-F12 code.\cite{usvyat2013} The fragment's SCF and canonical MP2 are carried out on the \textsc{Cryscor} side. Higher-level methods are run via an interface to the molecular code \textsc{Molpro}.\cite{molpro}
In this work we employ the canonical four-index integral \texttt{FCIDUMP} interface.\cite{Knowles89} In addition, we test an interface to our pilot implementation of density-fitted local correlation methods within the local integrated tensor framework (LITF)\cite{KatsManby2013} of the \textsc{Molpro} program package. It currently features local MP2, local distinguishable cluster with singles and doubles (local DCSD, LDCSD),\cite{Kats2013,katsAccurateThermochemistry2015} and local algebraic diagrammatic construction of second order (local ADC(2), LADC(2))\cite{Schirmer_ADC2,Trofimov_Schirmer_ADC2,dflcc2lr,flach2023} methods. For that the fragment's HF occupied orbitals are localized using the Pipek-Mezey localization procedure\cite{Pipek1989}, while the virtual space is represented by projected atomic orbitals (PAOs) restricted to the respective pair-specific domains.

\section{Computational parameters}

 Periodic DFT and HF calculations were performed with the \textsc{CRYSTAL17} program.\cite{Crystal17} Structural optimizations were carried out using the B3LYP functional\cite{becke1993} with the D3 dispersion correction\cite{Grimme2010} and Becke-Johnson damping.\cite{BJ2005}
SCF convergence was stabilized by damping the Fock matrix update to 80\% during the initial iterations. The DIIS acceleration procedure was activated only after the energy change dropped below $0.01$~hartree.

The initial structure of pristine phosphorene was obtained using a $12\times12$ k-mesh and a slightly modified POB-TZVP-rev2 basis set, where the d- and f-type functions were taken from the cc-pVTZ basis.\cite{Woon:93} In the defective structure calculations, however, this extended basis lead to numerical instabilities. Furthermore, subsequent CCSD(T) and EOM-CCSD calculations with this basis were prohibitively expensive except for the smallest fragments. Therefore, for all calculations beyond the pristine phosphorene structure optimization, we reverted to the standard pob-TZVP-rev2 basis set.\cite{VilelaOliveira2019}

The defect structure was optimized in a $4\times3$ supercell (47 atoms) using a $8\times8$ k-mesh, with one excess electron per cell to model the negatively charged vacancy. In addition, for a more accurate structural relaxation energy, the defect was re-optimized with a large $7\times6$ supercell (167 atoms) using a $6\times6$ k-mesh.
In all DFT calculations the default TOLINTEG tolerances of \textsc{Crystal17} (6 6 6 6 12) were used.

In the periodic HF calculations within the ADM, the integral tolerances were tightened significantly relative to the default values to minimize numerical errors in the embedding field. Specifically, the  \textsc{CRYSTAL} TOLINTEG thresholds were set to 10 10 10 30 100, and a $16\times16$ $k$-mesh was employed. Formally, the periodic HF calculation for the ADM could be performed in the primitive unit cell. In practice, however, \textsc{CRYSTAL} cannot evaluate the periodic Fock matrix $F^{\rm per}_{\mu'\nu'}$ for basis functions centered on displaced atoms when they appear too close to each other, which turned out to be the case within the primitive cell set up. Therefore, for the periodic HF step a $2\times2$ supercell was employed.

In the present density-fitted implementation of the ADM, the same auxiliary fitting basis is employed for both HF and post-HF two-electron integrals. Consequently, the auxiliary basis must provide adequate accuracy for both contributions. Test calculations show that the HF energy is considerably more sensitive to the choice of the auxiliary basis than the correlation energy, a behavior that has also been observed in other studies.\cite{Lambie2025}
Therefore, the main part (s-, p- and d-type orbitals) of our auxiliary basis were taken from the cc-pVTZ/JKFIT set,\cite{Weigend2002_JKFIT} which was optimized specifically for HF calculations. The $f$- and $g$-type orbitals were taken from the cc-pVTZ/MP2FIT auxiliary basis,\cite{Weigend2002_MP2FIT} which contains a richer set of high-angular-momentum functions. Such a combination was found to provide high accuracy in both HF and correlation energy differences evaluated with POB-TZVP-rev2 orbital basis.

The current local DCSD implementation within LITF is restricted to strong pairs, defined as pairs that share at least one atom in their respective orbital domains. The remaining pairs, denoted as weak pairs, are treated at the LMP2 level. For the ground-state amplitudes, the standard orbital domains extended by one bond connectivity (iext=1) were used.

Reference calculations for the phosphorus atom in its quartet ground state were also carried out using \textsc{Molpro} with ROHF,\cite{Roothaan1960_ROHF} RMP2,\cite{AMOS1991256} RDCSD,\cite{katsAccurateThermochemistry2015} and RCCSD(T).\cite{Watts93,Knowles93} To reduce the basis set superposition error, the phosphorus atom was surrounded by 11 ghost centers placed at the positions of neighboring atoms in the pristine phosphorene structure (for local methods it was 3 ghost atoms to mimic the iext=1 domains). All post-HF calculations used the frozen core approximation.

Finally, the periodic CIS optical band gap of pristine phosphorene was computed using the \textsc{Cryscor} CIS implementation described in Refs.~\onlinecite{Lorenz2011,Lorenz2012}.

\section{Structural specification of the ADM}\label{sec:struct}

\begin{figure}
\centering
\includegraphics[width=0.45\textwidth]{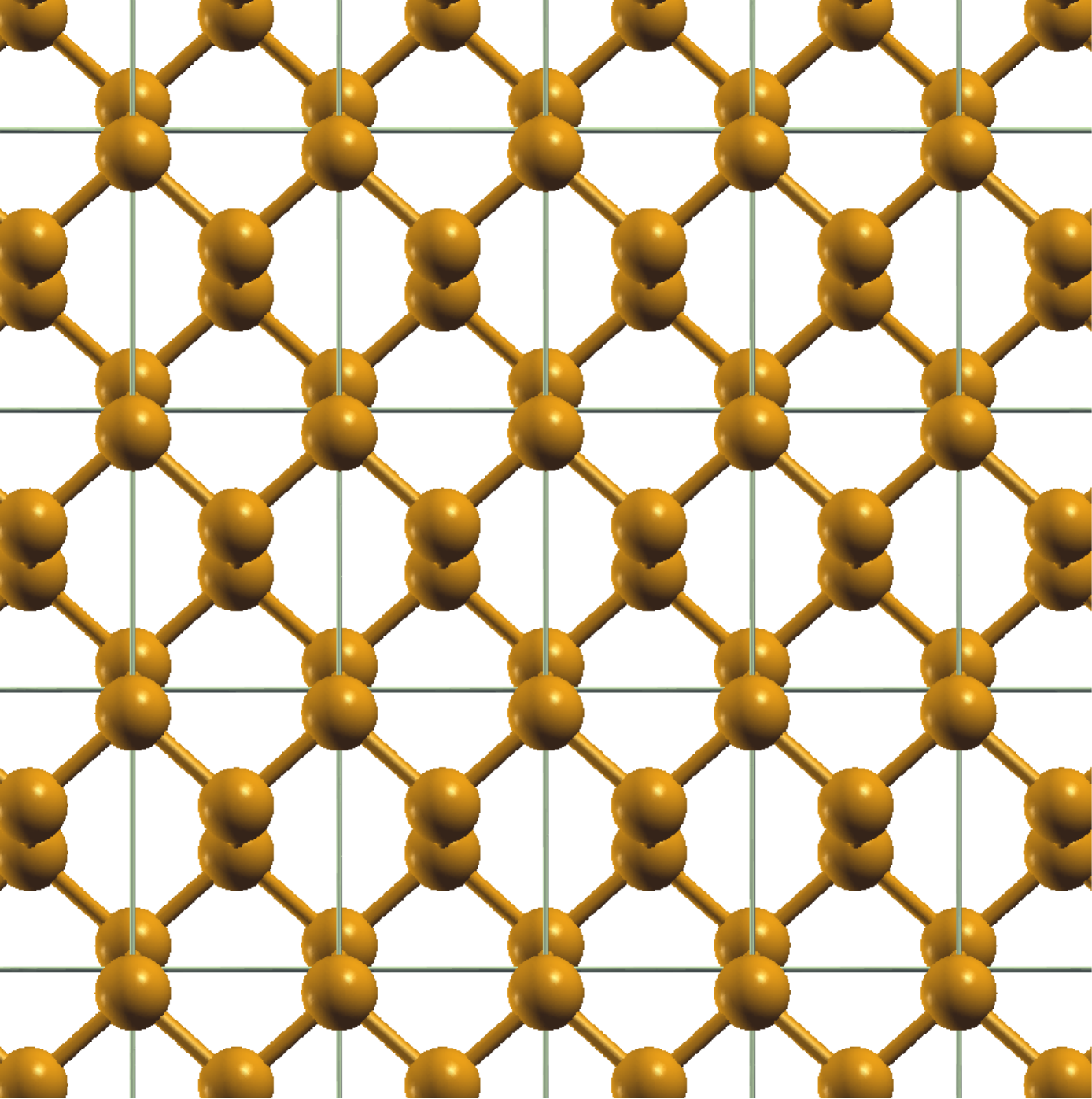}\\
(a)\\
\includegraphics[width=0.45\textwidth]{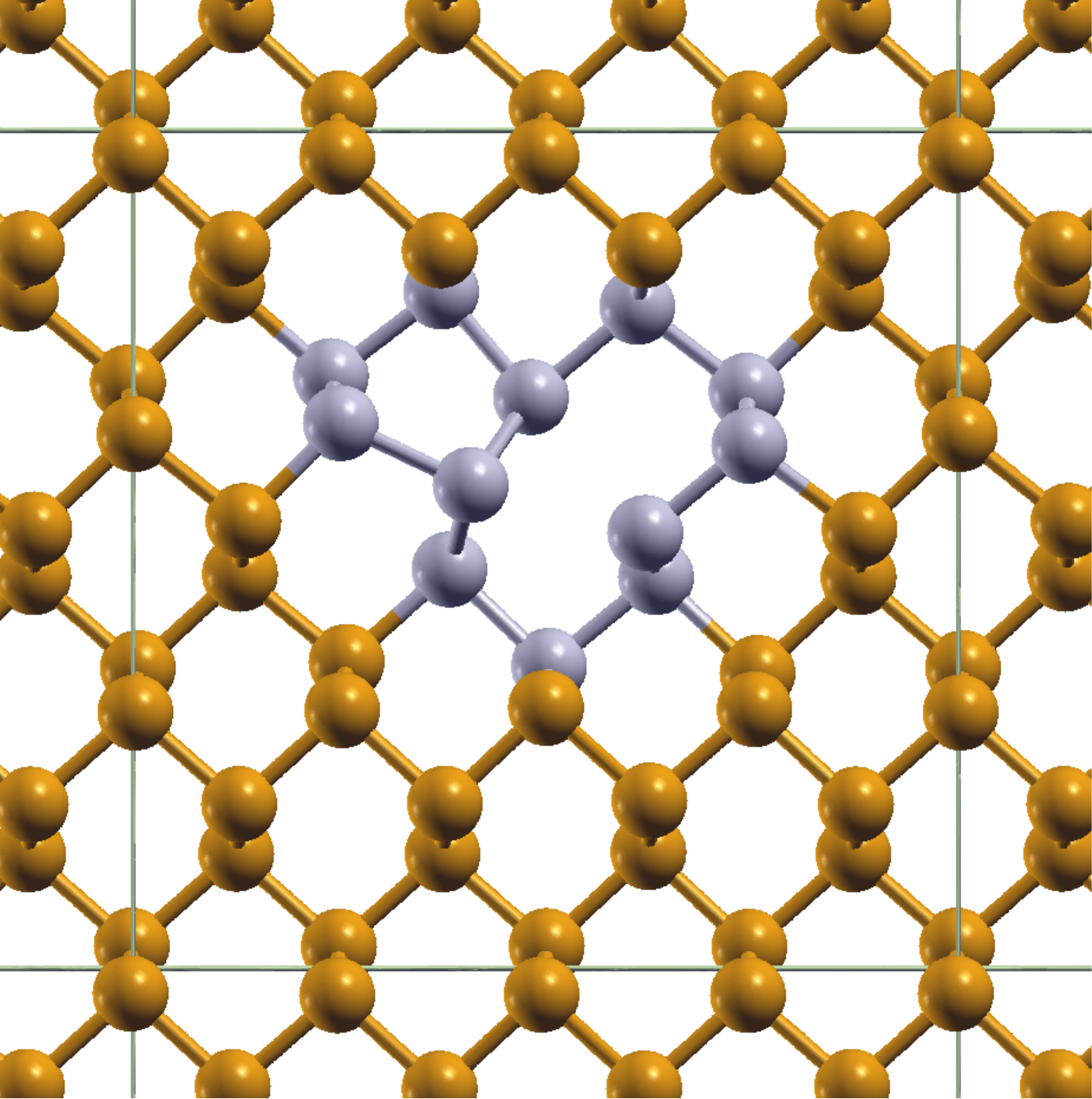}\\
(b)
\caption{Optimized structures of pristine phosphorene (a) and the negatively charged monovacancy in phosphorene within a $4\times3$ supercell (b). The structures of the pristine phosphorene (a) and of the negatively charged (5$|$9)-vacancy in phosphorene optimized within a 4$\times$3-supercell.  The atoms belonging to the reconstructed 5- and 9-membered rings are highlighted in gray.}\label{fig:per_phos}
\end{figure}

A B3LYP-D3 optimization of pristine phosphorene (Fig.~\ref{fig:per_phos}a) yields lattice parameters of $a=3.30\,\text{\AA}$ and $b=4.47\,\text{\AA}$ (the full structural specification is provided in the Supporting Information). Two reconstructed monovacancy structures are typically considered in phosphorene: the asymmetric (5$|$9) and the symmetric (55$|$66) configurations, denoted according to the number of atoms forming the rings. To compare these configurations energetically, we optimized the defect in a $4\times3$ supercell starting from both symmetric and distorted initial geometries. To probe the spin symmetry of the wavefunction of the negatively charged monovacancy, we considered both triplet and singlet spin states using broken-symmetry DFT. To prevent artificial trapping in a closed-shell solution in the singlet calculations, the initial guess orbitals were taken from preliminary triplet SCF runs allowing for possible spin polarization.

The optimized defect in the triplet state was found to be about 0.6 eV higher in energy than in the singlet state. When optimized starting from a distorted geometry, the singlet state converged to the (5$|$9) structure shown in Fig.~\ref{fig:per_phos}b. Furthermore, despite the spin-polarized initial guess,  at the (5$|$9) optimized geometry the wavefunction converged to a closed-shell solution.
Optimization starting from the symmetric structure resulted in a spin-polarized solution within the (55$|$66) geometry. However, this structure was slightly higher in energy (by about 0.04 eV), suggesting that it corresponds to a very shallow transition state between the (5$|$9) and (9$|$5) minima. These results are consistent with previous DFT studies of 
phosphorene monovacancies.\cite{Hu2015_PhosphoreneVacancies,Naik2020_PRMaterials_PhosphoreneDefects,Rijal2021_PhosphoreneVacancy}

\begin{figure*}
\centering
\begin{tabular}{ccc}
  \includegraphics[width=0.3\textwidth]{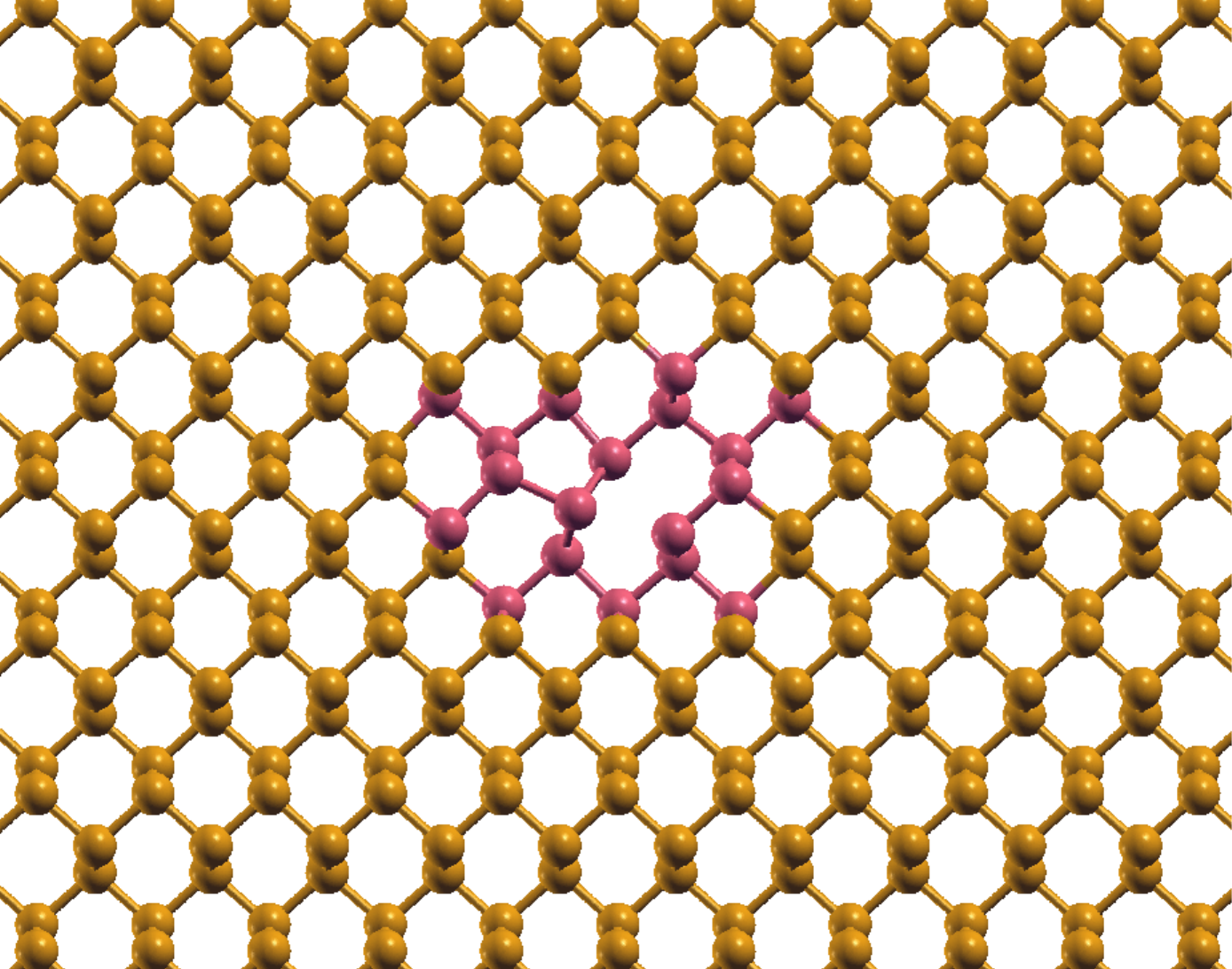}&
  \includegraphics[width=0.3\textwidth]{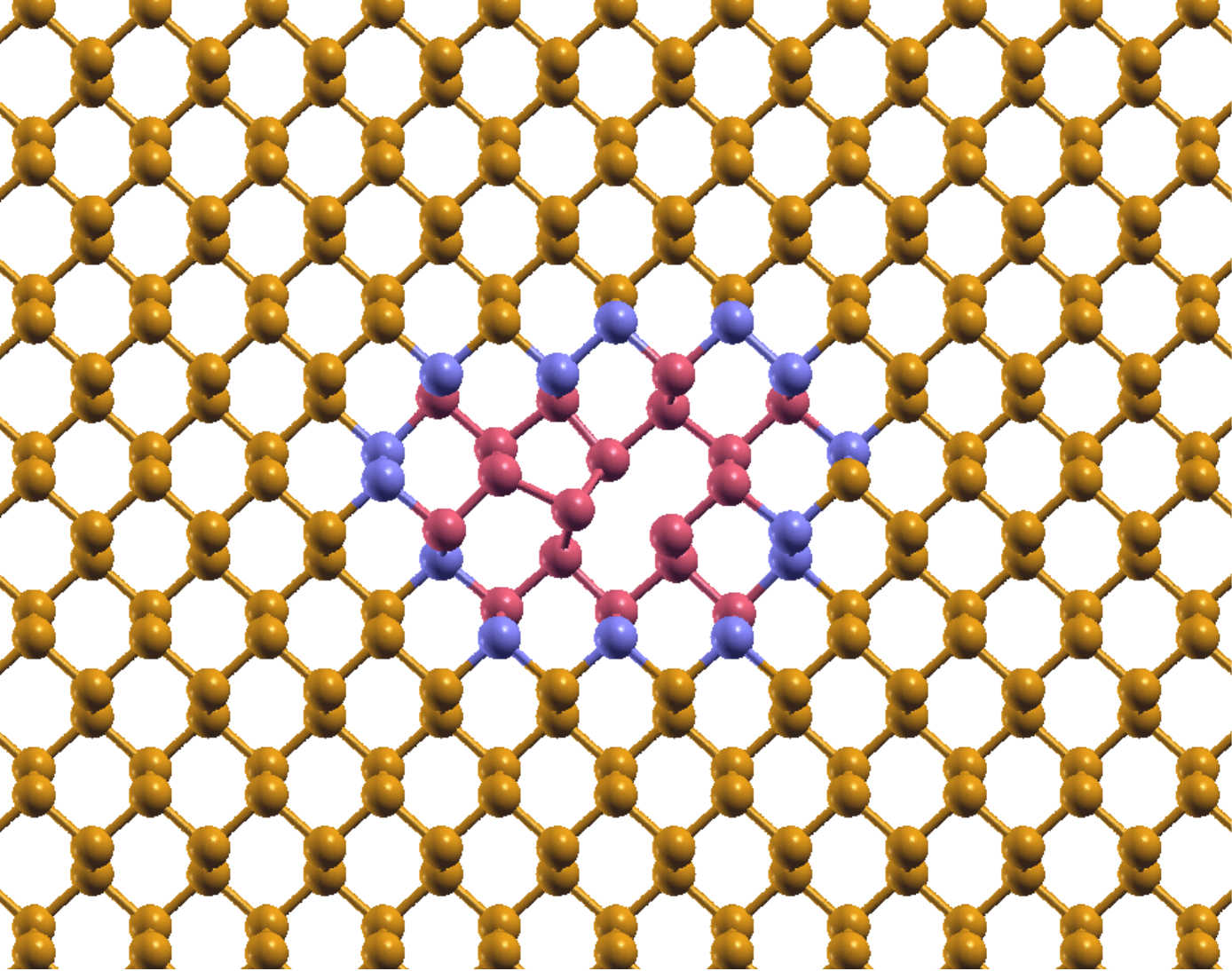}&
  \includegraphics[width=0.3\textwidth]{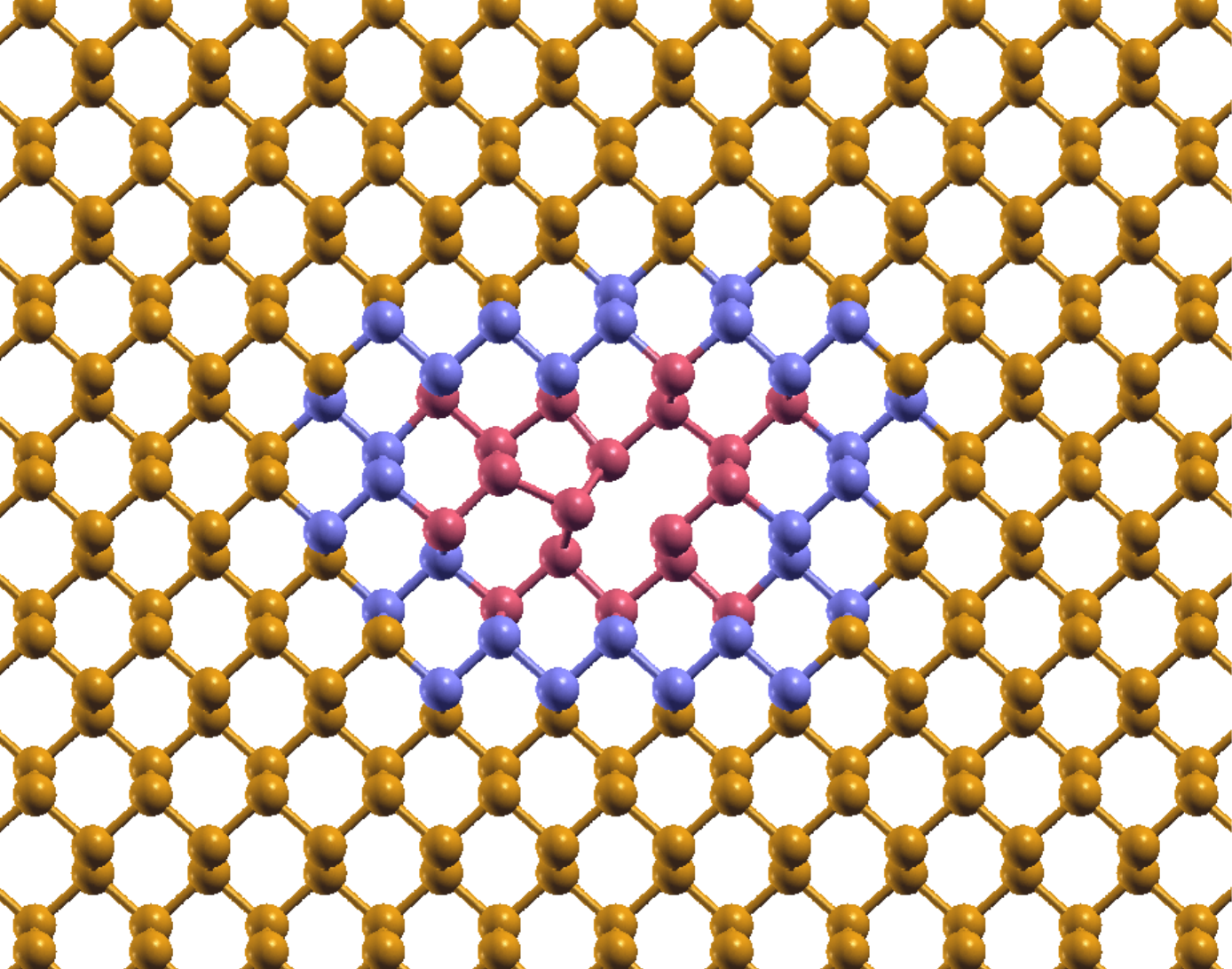}\\
  18 atoms & 32 atoms & 47 atoms\\
  \includegraphics[width=0.3\textwidth]{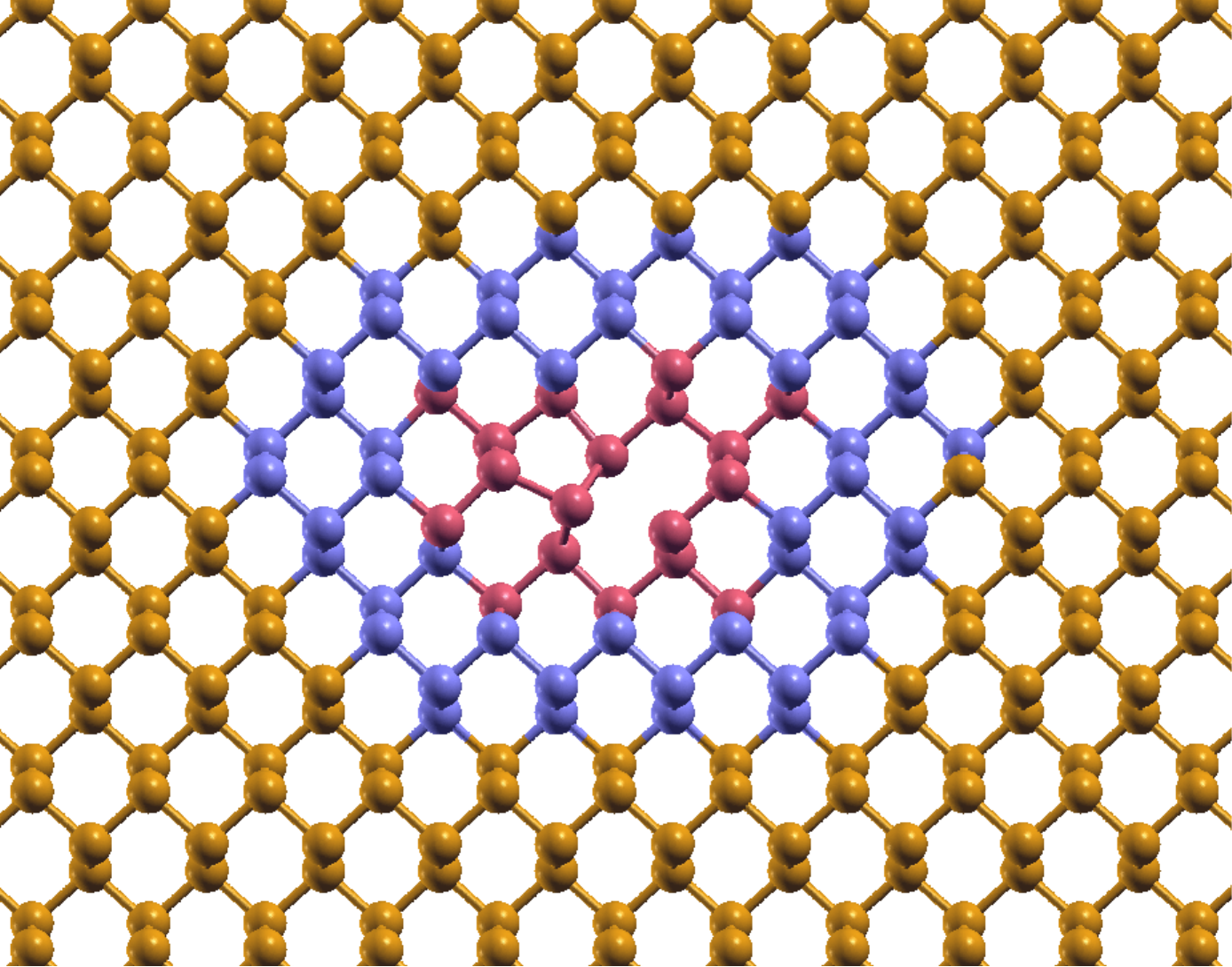}&
  \includegraphics[width=0.3\textwidth]{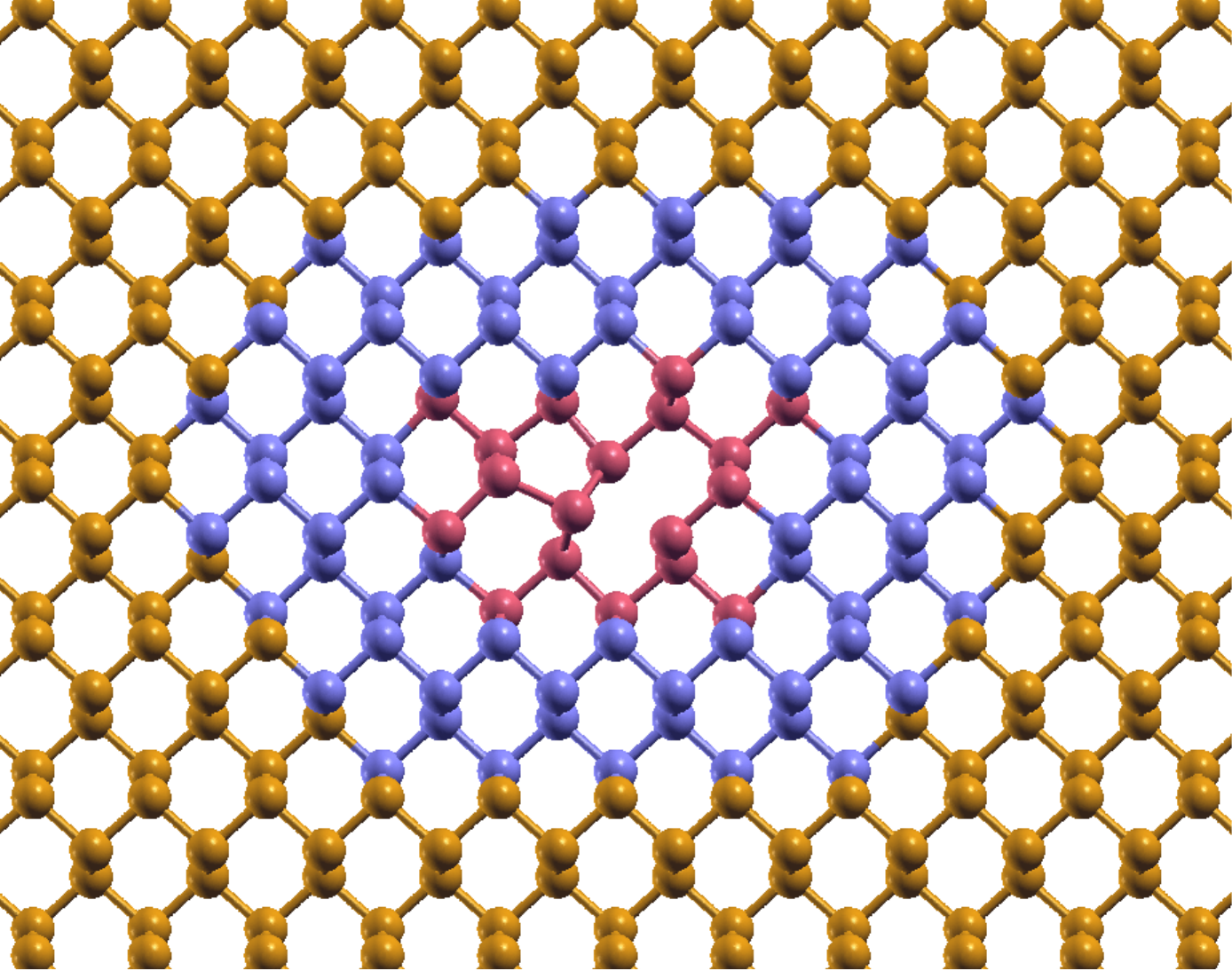}&
  \includegraphics[width=0.3\textwidth]{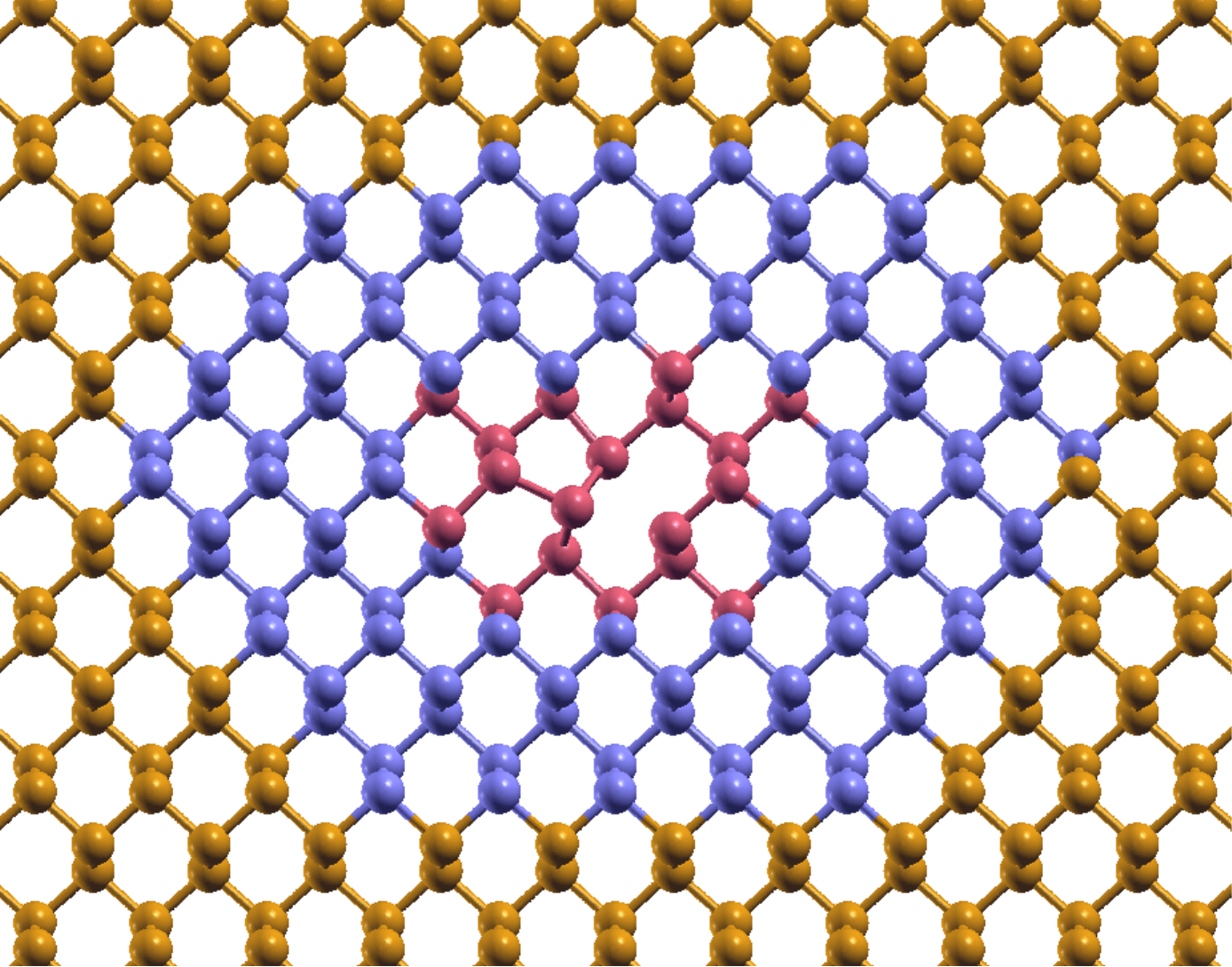}\\
  67 atoms & 88 atoms & 114 atoms\\
  &\includegraphics[width=0.3\textwidth]{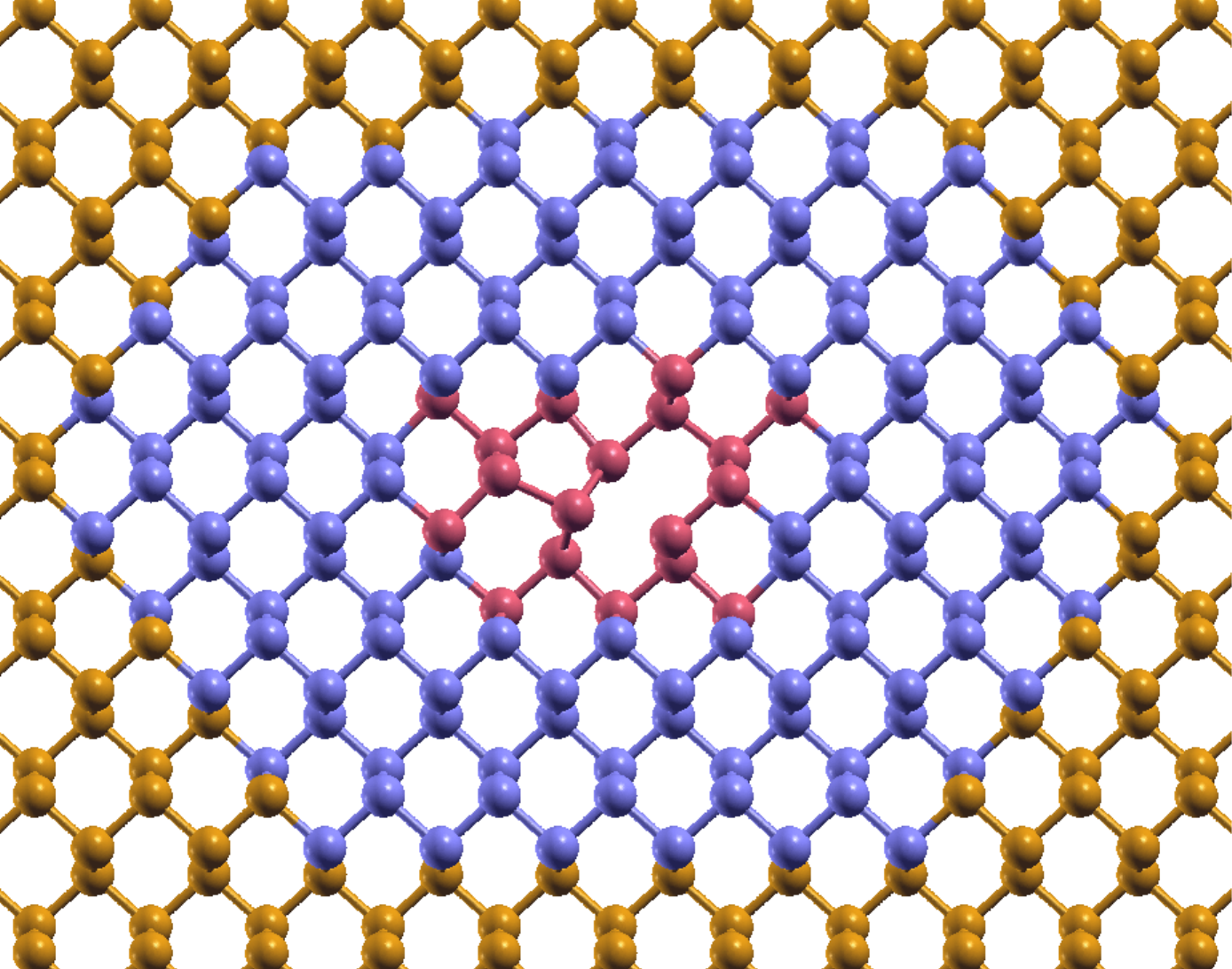}&\\
  & 141 atoms&
\end{tabular}
\caption{Fragments used in the ADM calculations. Atoms forming the defect, i.e., those whose positions are explicitly changed, are marked in pink. These atoms, and their counterparts from the pristine structure, explicitly enter the summations over $K'$ and $K$ in Eqs.~(\ref{eq:defect_h}) and (\ref{eq:E_nuc_defect}). Fragment atoms that remain in their positions but whose electrons are relaxed within the ADM are shown in blue. Atoms of the environment, whose electrons, represented by the respective WFs, are frozen and provide the embedding field, are shown in gold. The number of atoms in each fragment is indicated below the corresponding depiction.}\label{fig:frags}
\end{figure*}

Following these results, we focus on the minimum (5$|$9) structure for the subsequent ADM calculations. To enable the ADM calculations on fragments smaller than the full 47-atom supercell -- necessary for computationally demanding post-Hartree–Fock methods -- the system was reoptimized within the same $4\times3$ supercell while allowing only a subset of atoms to relax. Restricting relaxation to the 12 atoms that form the 5- and 9-membered rings (the gray atoms in Fig. \ref{fig:per_phos}b) leads to a collapse towards the (55$|$66) structure, indicating that the relaxed fragment must be further expanded. Including all atoms that move by at least 0.1~\AA\ adds six additional atoms, yielding an 18-atom fragment that correctly relaxes to the (5$|$9) configuration. This set of 18 atoms thus defines the minimal fragment used in the subsequent ADM calculations, as the latter must encompass all displaced centers (see Fig.~\ref{fig:frags}).

To assess the convergence with respect to fragment size, we consider a series of fragments containing up to 141 atoms shown in Fig.~\ref{fig:frags}. Each successive fragment is constructed by adding atoms within one bond connectivity from the previous fragment. The set of displaced atoms is kept fixed to the initially chosen 18 atoms.

For the defect formation energy $E_{\rm form}$ we employ its chemical definition, i.e. the electronic energy contribution to the energy of the model reaction
\begin{center}
  Phosphorene $\rightarrow$ Phosphorene with vacancy $+$ P-atom\\
\end{center}
It can be partitioned in several components:
\begin{eqnarray}
  E_{\rm form}&=&\Delta E^{\rm ADM}({\rm HF}) + \Delta E^{\rm ADM}({\rm corr})\nonumber\\
  &&+\Delta E_{\rm 18\uparrow relax},\label{eq:energy}
\end{eqnarray}
where $\Delta E^{\rm ADM}$ is the formation energy within the ADM model,
\begin{eqnarray}
  \Delta E^{\rm ADM}=E^{\rm ADM}_{\rm defect} + E_{\rm P} - E^{\rm ADM}_{\rm pristine},\label{eq:form_en}
\end{eqnarray}
with $E_{\rm P}$ being the energy of the phosphorus atom in its quartet ground state, and $\Delta E_{\rm 18\uparrow relax}$ -- the relaxation energy beyond the 18-atom fragment:
\begin{eqnarray}
  \Delta E_{\rm 18\uparrow relax}=E_{\rm relax} - E_{\rm relax\,18}.
\end{eqnarray}
Here we denote the energy of the fully optimized defect structure by $E_{\rm relax}$ and that of the structure with only 18 atoms allowed to relax by $E_{\rm relax\,18}$. With the $7\times6$ supercell and the B3LYP-D3 functional, $\Delta E_{\rm 18\uparrow relax}$ amounts to $-0.24$ eV.

We note that the definition (\ref{eq:energy}) corresponds to the reaction-energy formulation commonly used in quantum chemistry, in case the zero-point energies and thermal effects are neglected. It is equivalent to the standard solid-state expression for the defect formation energy\cite{Freysoldt2014Review}, provided that the phosphorus chemical potential is chosen as the energy of an isolated phosphorus atom. In contrast to periodic supercell approaches, charge and finite-size corrections\cite{Makov1995,Lany2008FiniteSizeErrors,Freysoldt2009ChargeCorrection,Komsa2013} are not required in the ADM, as the latter does not involve periodic replication of the defect or the introduction of a compensating background charge. Instead, the charged defect explicitly experiences the electrostatic potential of the non-defective crystal through the embedding. The ADM also provides direct access to the total energy of the charged species in its optimized structure, $E^{\rm ADM}_{\rm defect}$, eliminating the need for indirect and inherently approximate evaluations of the electron affinity based on defect-level energies.

Ideally, $\Delta E^{\rm ADM}$ should correspond to the thermodynamic limit, i.e., it should be converged with respect to the ADM fragment size or extrapolated to the infinite-fragment limit. The separation into HF and correlation components becomes useful here because the relatively inexpensive HF contribution, which contains most of the electrostatic and density-relaxation response to defect formation, may converge more slowly with fragment size than the more expensive correlation part. Furthermore, the correlation component itself can be partitioned into a lower-level contribution (e.g., MP2), which can be evaluated on large fragments, and a higher-level correction that is expected to converge even more rapidly.\cite{mullan21}

Technically, the ADM workflow for the present defect consists of an initial periodic calculation for pristine phosphorene, followed by orbital localization and evaluation of the electrostatic potential at the new positions of the 18 relaxed atoms. The latter enters the second term of Eq.~(\ref{eq:E_nuc_defect}). The defect is constructed by removing 19 atoms (one vacancy and 18 displaced atoms) and adding 18 atoms at their new relaxed positions. This defines the summation ranges over $K'$ and $K$ in Eqs.~(\ref{eq:defect_h}) and (\ref{eq:E_nuc_defect}).
Expansion of the fragment introduces additional basis orbitals $\ket{\mu'}$ and fragment Wannier functions $N_{\rm WFs}^{\rm frag}$, and thus additional electrons according to Eq.~(\ref{eq:n_el}). Finally, the SCF procedure, optionally followed by a post-Hartree-Fock treatment, is performed for the chosen fragment using the Hamiltonian in Eq.~(\ref{eq:defect_h}) and the projected basis functions defined in Eq.~(\ref{eq:new_AO}).  To account for the negatively charged defect, the electron count in the fragment calculations is increased by one electron by setting  $\Delta_{\rm el}=1$ in Eq.~(\ref{eq:n_el}).

Importantly, to evaluate $\Delta E^{\rm ADM}$ consistently, the set of explicitly manipulated atoms entering the sums over $K'$ and $K$ in Eq.~(\ref{eq:E_nuc_defect}) must correspond one-to-one between the defect and pristine systems.  This implies that in the ADM calculation on the pristine structure the atoms, corresponding to the removed or displaced atoms of the defect, must be formally marked as both removed ($K$) and added ($K'$), so that they explicitly enter the effective nuclear-energy term in Eq.~(\ref{eq:E_nuc_defect}). Obviously the fragments have also to be consistent between the defect and pristine ADM.


\section{Results and Discussion}

\subsection{Defect Formation Energy}\label{sec:form_en}

We begin by analyzing the Hartree-Fock contribution $\Delta E^{\rm ADM}({\rm HF})$ to the defect formation energy, shown in Fig.~\ref{fig:En_form} (the numerical values are provided in the Supplementary Material). 
The smallest 18-atom fragment is clearly inadequate at the HF level. This is not surprising, as for this fragment some bonding Wannier functions of the environment remain effectively anchored to the original atomic positions left behind by the displaced atoms. But even for larger fragments, the HF contribution remains far from converged, and realistic formation energies are only obtained for fragments containing 67 atoms or more. A deviation within 0.1~eV from the extrapolated value is reached only for the largest fragment considered (141 atoms). 

Despite these sizable finite-size errors for small and medium fragments, the convergence behavior is remarkably systematic. When plotted as a function of $1/N_{\rm at}^{2.8}$, the decay of $\Delta E^{\rm ADM}({\rm HF})$ becomes nearly linear over a wide range of fragment sizes, enabling a reliable extrapolation to the thermodynamic limit. Notably, this convergence is significantly faster than the $1/N_{\rm at}$ behavior expected for purely electrostatic finite-size effects of a charged defect. This observation suggests that the dominant source of the slow convergence is not long-range electrostatics, but rather the redistribution of the electronic density along the bonds induced by the structural rearrangement around the vacancy.

\begin{figure}
\centering
\includegraphics[width=0.47\textwidth]{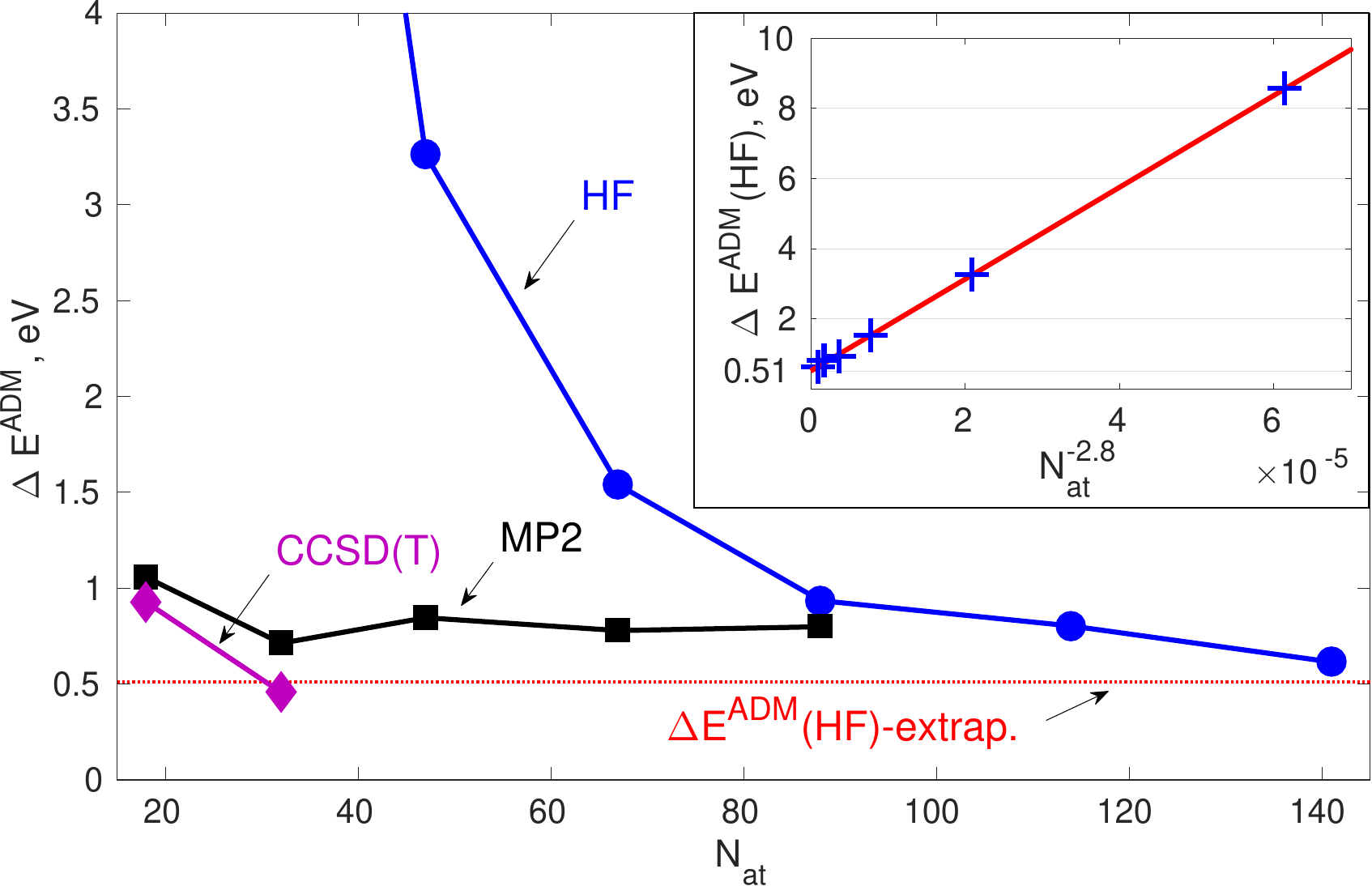}
\caption{HF and correlation components of the defect formation energy as functions of the fragment size $N_{\rm at}$. The inset shows the HF formation energy plotted as a function of $1/N_{\rm at}^{2.8}$, along with the linear fit to extrapolate it to the thermodynamic limit.}
\label{fig:En_form}
\end{figure}

Although $\Delta E^{\rm ADM}({\rm HF})$ requires relatively large fragments for convergence, the computational cost of the density-fitted HF calculation remains modest, allowing the treatment of such fragment sizes. In contrast, the correlation contribution, which is computationally significantly more demanding, converges much more rapidly. Even the smallest fragment already yields meaningful values for the correlation energy, and for the second (32-atom) fragment, $\Delta E^{\rm ADM}({\rm MP2})$ is converged to within 0.1~eV.

Coupled-cluster calculations at the CCSD(T) level are too expensive to be performed beyond the 32-atom fragment. Unfortunately, for polarizable systems the convergence behavior of MP2 may differ from that of CCSD(T). In particular, long-range dispersion interactions in the latter can be effectively screened by many-body effects, which are not captured at the MP2 level.\cite{schutz2017,usvyat18} As a result, MP2 tends to overestimate long-range correlation contributions in such systems. This effect likely explains the more pronounced drop in $\Delta E^{\rm ADM}({\rm CCSD(T)})$ compared to MP2 when increasing the fragment size from 18 to 32 atoms. Nevertheless, since even the unscreened MP2 contribution is essentially converged at the 32-atom fragment, the corresponding CCSD(T) result can also be expected to be close to convergence.

We are now in a position to provide a CCSD(T)/POB-TZVP-rev2 estimate of the formation energy of a negatively charged monovacancy in phosphorene. Its Hartree-Fock component, $\Delta E^{\rm ADM}({\rm HF})$, extrapolated to the thermodynamic limit, amounts to 0.51~eV. The correlation contribution, $\Delta E^{\rm ADM}({\rm corr})$, is evaluated as the MP2 result obtained with the 89-atom fragment (0.803~eV), combined with a CCSD(T) correction to MP2 computed with the 32-atom fragment (-0.256~eV), yielding a total correlation contribution of 0.543~eV. Finally, the relaxation correction, $\Delta E_{\rm 18\uparrow \rm relax}$, calculated on a 167-atom supercell, is -0.24~eV (see Sec.~\ref{sec:struct}). 
Combining these contributions gives a final estimate of the defect formation energy of $E_{\rm form} = 0.81$~eV$\approx 18.7$~kcal/mol.
This moderate formation energy indicates that vacancy formation in phosphorene is  energetically accessible, suggesting that such defects can form under realistic conditions in appreciable concentrations, which is in agreement with experimental studies.\cite{Kiraly2017_BPvacancies,Fang2022}

The present value lies at the lower end of the range obtained from DFT supercell calculations for neutral vacancies (0.9--1.7~eV).\cite{Hu2015_PhosphoreneVacancies,Naik2020_PRMaterials_PhosphoreneDefects,Rijal2021_PhosphoreneVacancy} For the charged vacancy this value was corrected by adding the electron affinity, obtained as a Kohn-Sham- or quasi-particle energy level associated with the defect relative to the valence band maximum. Depending on the theoretical level (DFT or DFT+GW) and the functional employed, the electron affinity has been reported in the range of 0.6--1.0~eV.\cite{Naik2020_PRMaterials_PhosphoreneDefects,Rijal2021_PhosphoreneVacancy}
In contrast, the ADM enables a direct evaluation of formation energies for both neutral and charged defects on the same footing, as differences between total energies, while also allowing the use of high-level quantum-chemical methods for their accurate determination.

Conservative error bars associated with the remaining finite-size effects in each of the three components, $\Delta E^{\rm ADM}({\rm HF})$, $\Delta E^{\rm ADM}({\rm corr})$, and $\Delta E_{\rm 18\uparrow \rm relax}$, are expected to be below 0.1~eV. The dominant source of uncertainty is likely the basis-set incompleteness. Although the POB-TZVP-rev2 basis performs reasonably well in periodic DFT, it does not guarantee near basis-set convergence in our calculations, particularly for the correlation energy. In principle, the ADM allows the use of different, potentially much larger molecular basis sets for the fragment calculations. However, this capability is not yet implemented in the present code, and a systematic assessment of the basis-set incompleteness error is deferred to future work.

\subsection{Local correlation treatment}

With canonical high-level methods (beyond MP2), the fragment sizes required to observe a clear convergence pattern may become prohibitively large, so that convergence can often only be assessed indirectly via lower-level approaches, like MP2. The underlying reason is the steep scaling of the computational cost of accurate quantum-chemical models, which can, however, be mitigated by employing local correlation schemes.\cite{Schuetz2001_LCCSD} The resulting reduced scaling of the local approach enables quantum-chemical calculations on substantially larger systems. In the present context, this allows the treatment of much larger fragments and thus a direct assessment of convergence with respect to fragment size. In this work, we employ our pilot implementation of the local distinguishable cluster method interfaced with the ADM framework.

The correlation contributions $\Delta E^{\rm ADM}({\rm corr})$ to the defect formation energies, computed using both canonical and local methods, are compiled in Table~\ref{tab:local}. We first note that the canonical DCSD results are in good agreement with those of CCSD(T). In particular, the smaller magnitude and faster decay of $\Delta E^{\rm ADM}({\rm CCSD(T)})$ compared to MP2 are also reproduced at the DCSD level.

\begin{table}[h]
  \centering
  \caption{Correlation contribution $\Delta E^{\rm ADM}({\rm corr})$ to the defect formation energy calculated for several fragments using canonical and local correlation methods. For the local methods, the correlation energy is further partitioned into strong- and weak-pair contributions, depending on the pair types included in the pair-energy summation. Since the current implementation of LDCSD is restricted to strong pairs only, its weak-pair contribution is equal to that of LMP2. The energies are in eV.}
    \begin{tabular}{lcccccc}
    \hline\hline
    &\multicolumn{6}{c}{$N_{\rm at}$}\\
    &&18 && 32 &&47 \\
    \hline
    
    \multicolumn{7}{c}{   Canonical treatment}\\
   MP2           &~~ & 1.057 &~~ & ~0.713 &~~ & ~0.846 \\
   DCSD          & & 0.862 & & ~0.420  &&       \\
   CCSD(T)       &  & 0.927 & & ~0.457 & &       \\
    \hline
    \multicolumn{7}{c}{    Local treatment}\\
   LMP2          & & 1.074 & & ~0.713 & & ~0.838 \\
   LDCSD         & & 1.122 & & ~0.674 & & ~0.805 \\
    \hline   
   LMP2 strong   & & 0.174 & &-0.270 & & -0.254 \\
   LMP2 weak     & & 0.900 & & ~0.983 & &  ~1.092 \\
   LDCSD strong  & & 0.222  &&-0.309 & &  -0.287 \\
    \hline \hline
    \end{tabular}
    \label{tab:local}
\end{table}

The LMP2 results are in close agreement with canonical MP2, indicating that the error introduced by the local domain approximation is small. In contrast, the LDCSD results deviate significantly from canonical DCSD and instead remain close to LMP2.  To rationalize this behavior, we decompose the correlation energy into contributions from short-range (strong pairs) and long-range (weak pairs). We note that the present LDCSD implementation is restricted to strong pairs only, following the standard pair approximation employed in earlier local correlation schemes.\cite{Schuetz2001_LCCSD} As a result, the long-range correlation contribution in LDCSD is effectively described at the LMP2 level.

Analysis of the short- and long-range contributions to $\Delta E^{\rm ADM}({\rm corr})$ shows that its dominant part arises from the long-range component, i.e., from the difference in dispersive interactions between the defective and pristine systems. This observation explains why CCSD(T) and DCSD, which account for Coulomb screening of dispersion interactions, yield smaller values of $\Delta E^{\rm ADM}({\rm corr})$ compared to MP2, where dispersion is effectively unscreened. Indeed, the confinement of dispersion interactions due to screening is  known to be pronounced in polarizable systems in general and has been observed in black phosphorus or phosphorene.\cite{schutz2017,usvyat18}  
The dominance of the long-range contribution also supports our, albeit indirect, assessment that $\Delta E^{\rm ADM}({\rm CCSD(T)})$ converges rapidly with fragment size, as the long-range correlation is expected to decay faster at the CCSD(T) level than at the MP2 level.

Under these circumstances, if the weak-pair contribution in LDCSD remains described at the MP2 level, the accuracy of $\Delta E^{\rm ADM}({\rm LDCSD})$ is compromised. This highlights the importance of a hierarchical pair approximation,\cite{MasurUsvyatSchutz_rCCD3,schutz2014,Schwilk2015,Schwilk17} in which weak pairs are treated at an approximate, yet still coupled-cluster-based level. Such an approach enables an accurate and at the same time computationally affordable description of dispersion. Implementation of local correlation methods within the ADM employing a proper pair treatment is currently underway.

As for the short-range correlation, which is treated at the true LDCSD level, its small but non-negligible contribution to $\Delta E^{\rm ADM}({\rm LDCSD})$ converges rapidly with fragment size, as evident from Table~\ref{tab:local}. Together with the indirect indication of the fast convergence of the long-range component, this supports the conclusion that the total high-level correlation contribution reported in Sect.~\ref{sec:form_en} is well converged.

\subsection{Excitation Energy}

Finally, we examine the excitation energies associated with the defect.
The ADM enables a high-level quantum-chemical treatment of excited states localized on defects. Such a treatment is particularly significant for solid-state applications, where excited states are still predominantly described at the level of the one-particle Kohn-Sham band structures.

\begin{figure}
\centering
\includegraphics[width=0.47\textwidth]{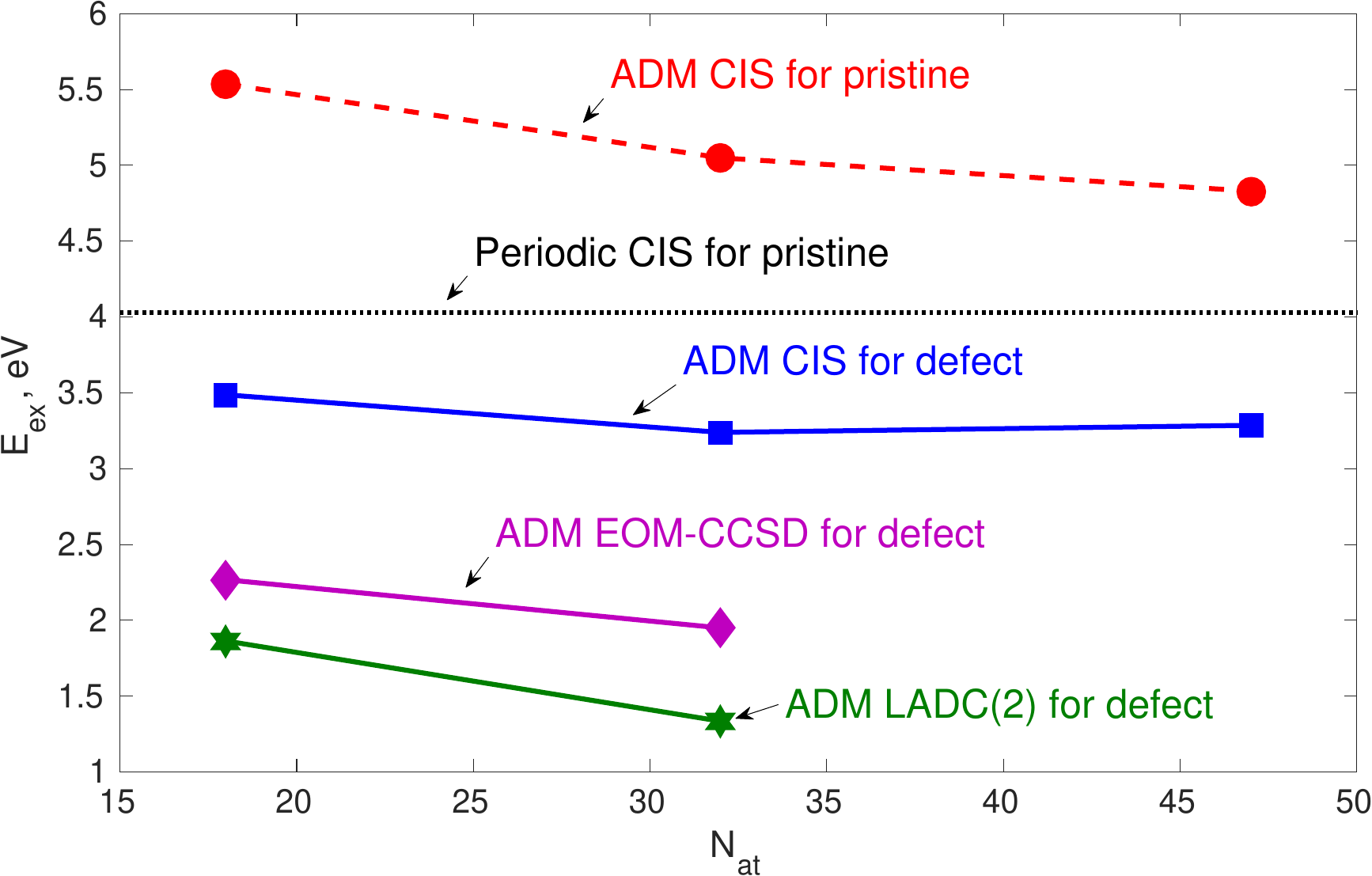}
\caption{First excitation energy of the defect calculated using CIS, LADC(2), and EOM-CCSD methods, as well as the CIS excitation energy of pristine phosphorene computed both fully periodically and within the ADM.}
\label{fig:En_ex}
\end{figure}

We note that the ADM is effective for states that lie within the band gap, i.e., whose excitation energies are lower than the lowest excitation energy of the host solid. Otherwise, in the absence of constraints on the excited-state wavefunction, expansion of the fragment would lead to collapse into bulk-like excited states. We illustrate this behavior using pristine phosphorene, treated at the CIS level both fully periodically and within the ADM (see Fig.~\ref{fig:En_ex}). Periodic CIS, as implemented in \textsc{Cryscor}, allows one to compute the optical band gap at the $\Gamma$-point.\cite{Lorenz2011,Lorenz2012} Although CIS significantly overestimates the band gap, it provides a useful reference for the relative energetic position of localized excited states.

When pristine phosphorene is treated within the ADM, the lowest excitation confined to the fragment lies well above the band gap. Upon increasing the fragment size, the excitation energy gradually decreases. In the thermodynamic limit, the ADM excited state is expected to collapse into a delocalized bulk state, with the excitation energy dropping to the periodic band gap value.

In contrast, for a defect that supports localized excited states, the ADM becomes highly effective. As seen in Fig.~\ref{fig:En_ex}, the CIS excitation energy of the defect lies below the CIS band gap of phosphorene and converges already for a 32-atom fragment. As the value obtained with the 47-atom fragment is even slightly higher than that for the 32-atom fragment (differing by less than 0.05~eV), it is reasonable to assume that convergence has essentially been reached.

CIS itself is not sufficient for quantitatively accurate excitation energies; therefore, we also employed local ADC(2) and canonical EOM-CCSD using respective interfaces. EOM-CCSD is a highly accurate method, with typical errors in excitation energies on the order of 0.1~eV for states dominated by single excitations (excluding possible basis-set incompleteness effects).\cite{Loos2018_JCTC_Benchmark,Loos2020_JCTC_Benchmark} ADC(2) is a second-order perturbative approach which, although not a direct approximation to EOM-CCSD, can be regarded as a computationally more affordable but less accurate alternative.

Unfortunately, both canonical EOM-CCSD and local ADC(2) treatments turned out to be computationally unfeasible for fragments larger than 32 atoms. However, for the 18- and 32-atom fragments, the EOM-CCSD excitation energies are virtually parallel to the CIS ones. This suggests that the 32-atom EOM-CCSD result is also likely to be close to convergence. The resulting value of 1.95~eV is therefore taken as our best estimate for the excitation energy associated with the negatively charged vacancy in phosphorene. While this value is obtained from a high-level quantum-chemical treatment, it may still be affected by basis-set incompleteness.

Interestingly, the LADC(2) excitation energies are noticeably lower (by as much as 0.6~eV for the 32-atom fragment) than those of EOM-CCSD and exhibit a stronger dependence on fragment size. Two factors may contribute to this deviation. First, as a perturbative method, ADC(2) may suffer from deficiencies similar to those of ground-state MP2, in particular the lack of Coulomb screening, which can lead to an overestimation of long-range correlation effects. Second, in the local treatment the virtual space used for ground-state LMP2 amplitudes is restricted to smaller domains than the excited-state amplitudes, for which the domains are extended to properly describe the character of the excited-state wavefunction. This imbalance in the local approximations between ground and excited states can effectively lead to an artificial lowering of the excitation energy.

\section{Conclusions and outlook}

In this work, we demonstrate that the ADM represents a highly promising paradigm for studying defects in periodic systems. It provides a framework with several key advantages over the standard supercell approach. Most notably, it treats a {\it single} defect rather than a periodically repeated array of defects, eliminating spurious interactions between defect images and removing the need for charge-correction schemes in the case of charged defects.
Furthermore, the ADM reduces a periodic problem to an effective molecular-like calculation on a fragment of atoms, with a modified Hamiltonian and an atomic-orbital-like basis set. Beyond the associated reduction in computational cost, this brings along conceptual advantages of molecular quantum chemistry, such as systematically improvable high-level methods, a direct access to charged, high-spin or excited defect states, the possibility of multireference treatments for strongly correlated defects, etc.

A conceptual limitation of the ADM in its current formulation is its inapplicability to defects in conducting systems. This stems from the environment–fragment partitioning based on localized occupied orbitals, which cannot be well localized in metals, rendering such partitioning as physically questionable.
For non-metallic systems, challenges arise for defects whose states extend deeply into the environment and induce long-range electronic or structural rearrangements, leading to slow convergence with respect to fragment size. Supercell approaches, however, may be similarly, if not more, affected by such defect delocalization.

Our calculations for a vacancy in phosphorene demonstrate the efficiency of the ADM on a realistic and challenging example, involving both ground and excited states of a charged defect in a highly polarizable semiconductor. Using the ADM allowed us to provide high-level benchmark estimates for both the formation energy and the lowest excitation energy. At the same time, the calculations revealed several important aspects relevant for practical applications of the ADM that remain to be addressed in future developments.

First, the benefits of high-level electronic-structure methods may be partially offset by limitations of the employed basis sets. Periodic calculations are known to become numerically unstable with large atomic-orbital basis sets due to the onset of quasi-linear dependencies among the basis functions. The ADM conceptually allows one to redefine the fragment basis and thus employ systematically improvable molecular basis-set hierarchies for the fragment part of the calculation. The implementation of this feature is currently underway. Another promising avenue in this respect is the extension of the ADM to a transcorrelated framework.\cite{Christlmaier23} Recently, some of us have demonstrated for defects in silicon that a transcorrelated treatment of the fragment combined with a mean-field transcorrelated embedding allows for a substantial acceleration of basis-set convergence for defect formation energies.\cite{simula2025transcorrelated}

Second, at present the structural optimizations of defects rely on DFT supercell calculations. In this respect, it would be important to generalize the ADM framework to density functional theory, including the implementation of nuclear gradients. This would not only enable the treatment of larger fragments using computationally inexpensive non-hybrid functionals, but also allow for geometry optimization directly within the ADM. In addition, the availability of nuclear gradients for high-level quantum-chemical methods would facilitate structural refinement in the immediate vicinity of the defect. This is particularly important for strongly correlated defects, for which DFT equilibrium structures may be unreliable.

In the phosphorene example, relatively large fragments were required to converge the HF contribution to the formation energy. Therefore, accelerating convergence with respect to fragment size is also highly desirable. A possible route for this is to permit approximate relaxation of the environment due to the defect formation and incorporate its response onto the fragment, for example following the density matrix embedding theory (DMET)\cite{PhysRevLett.109.186404,Knizia2013,Cui2020,Pham2020} framework.

In contrast, the correlation contribution in the present system converges rather rapidly with fragment size. This behavior, however, may not be general, particularly in systems where dispersion interactions play a significant role (e.g., adsorbates on surfaces).
To enable the treatment of large fragments at the correlated level, it will be important to either implement lower-scaling approaches, such as SVD-decomposed\cite{lesiuk2019efficient,Rickert2025} or local correlation methods,\cite{Ma18} within the ADM or interface the ADM with existing implementations. As the present results also demonstrate, an accurate treatment of weak pairs within the local coupled-cluster framework\cite{schutz2014,Schwilk2015} is essential.

\section*{Supplementary Material}

The Supplementary Material provides the raw computational data, the detailed specifications of the pristine and defect structures considered in this work.

\section*{Acknowledgements}
C.R. gratefully acknowledges Studienstiftung des deutschen Volkes for a Master's scholarship and Tiemann Stiftung for financial support of research visits  to the Max Planck Institute for Solid State Research.


%

\end{document}